\documentstyle[11pt,newpasp,twoside]{article}
\def\edcomment#1{\iffalse\marginpar{\raggedright\sl#1\/}\else\relax\fi}
\reversemarginpar

\begin{document}
\title{A molecular gas phase in the Cold Neutral Medium?}
 \author{Padeli P. Papadopoulos}
\affil{Dept. of Physics and Astronomy, University College London,
 Gower Street, London, WC1E 6BT, UK}
%\author{Ima Co-Author}
%\affil{The Name of My Institution, The Full Address of My Institution}

\begin{abstract}
I  examine  the  possibility  that  the Cold  Neutral  Medium  of  the
interstellar medium  in galaxies contains  a molecular gas  phase that
may represent a  significant and even the dominant  amount of its mass
in metal-poor  regions. In spiral  galaxies such regions are  found at
large  galactocentric  distances  where  diffuse  H$_2$  gas  will  be
untraceable through  its feeble $  ^{12}$CO J=1--0 emission,  the very
lack  of it  being also  responsible its  higher  kinetic temperatures
($\rm T_k\sim 60-100$ K).  Sensitive sub-mm imaging of spiral galaxies
has  demonstrated the  existence  of  dust well  inside  their HI  gas
distribution  while recent  observational work  suggests a  high H$_2$
formation rate from HI association onto grains. The latter is indeed a
critical  unknown and  a  high  value can  easily  compensate for  the
reduction   of  the   available  grain   surface  in   the  metal-poor
environments giving  rise to a  CO-poor, diffuse H$_2$ phase  that may
then  contribute  significantly  to  the  mass  and  pressure  of  the
interstellar medium in such environments.
\end{abstract}

\section{Introduction}

The use of CO rotational lines to detect the bulk of H$_2$ gas mass in
galaxies and deduce its  physical properties is now a well-established
technique (e.g.   Dickman, Snell, \& Schloerb 1986;  Young \& Scoville
1991) employed  successfully  in the  local  Universe (e.g.   Sanders,
Scoville  \& Soifer  1991; Downes  \& Solomon  1998) as  well  as high
redshifts  (e.g.  Brown  \& vanden  Bout  1991; Omont  et al.   1996).
Nevertheless early  studies have indicated serious  limitations of the
method in  tracing metal-poor H$_2$  gas (e.g. Maloney \&  Black 1988;
Israel  1997).    These  become  particularly   severe  in  metal-poor
environments  where the  reduced dust-shielding  allows UV  photons to
dissociate the  CO molecule  while leaving the  largely self-shielding
H$_2$ intact.  This is now corroborated by firm observational evidence
for the metal-poor outer parts of typical spirals (Nakai \& Kuno 1995;
Arimoto  et  al.   1996),  in  globally metal-poor  objects  like  the
Magellanic Clouds  and Magellanic  Irregulars (Israel 1997;  Madden et
al.  1997),  and blue  compact dwarf galaxies  (Barone et  al.  2000).
Interestingly  in  many  such  environments pressure  ``probes''  like
diffuse CO-bright  molecular clouds (Heyer, Carpenter,  \& Snell 2001)
or HII regions (Rudolph et al.  1996; Elmegreen, \& Hunter 2000) imply
unexpectedly  high pressures  which  then suggest  the  presence of  a
pressure agent other than HI or CO-bright H$_2$ gas.

\section{H$_{\bf 2}$ formation rate: its role in metal-poor environments}

The dominant formation mechanism of H$_2$ in interstellar environments
is by HI association onto dust grains. The dust distribution according
to recent  sensitive maps  of its sub-mm  emission extends  beyond the
optical  and  well into  the  HI disk  of  a  typical spiral  (Nelson,
Zaritsky, \& Cutri  1998; Xylouris et al.  1999),  and possibly out to
$\sim  2\rm   R_{25}$  (Thomas  et   al.   2002;  Guillandre   et  al.
2001). {\it Hence the two ingredients for H$_2$ formation can be found
to the outermost and metal-poor  parts of typical spiral disks.}  From
Federman, Glassgold, \& Kwan (1979) and an exponent of m=1.5 for their
H$_2$  self-shielding function it  can be  shown that  an HI  cloud of
radius  $\rm   R_{\rm  cl}$  (pc),  average  volume   density  n,  and
temperature  $\rm  T_k$, illuminated  by  a  FUV  field strength  $\rm
G_{\circ}$  ($\rm G_{\circ}$(solar)=1)  starts turning  molecular when
$\rm S>1$, where

\begin{equation}
\rm S=\frac{\rm n}{\rm n_{\rm min}}=\rm R_{\rm cl} ^{2/5}\left(\frac{\rm n}{65\rm 
cm^{-3}}\right) \left(\frac{\rm T}{100\rm K}\right)^{3/10}\left(
\frac{\mu \rm z}{\rm G_{\circ }}\right)^{3/5}.
\end{equation}

\noindent
The density $\rm n_{\rm min}$ is the minimum required for the onset of
the $\rm H\rightarrow H_2$ transition  and $\rm z= Z/Z_{\odot}$ is the
ambient metallicity normalized to its solar value. We assumed that the
H$_2$ formation rate of $\rm R_{f}=S_{f} T^{1/2}$ scales linearly with
total grain surface and thus metallicity, and included the factor $\mu
= \rm S_{f}/S^{(o)}  _{f}$, to parametrize for a  value other than the
canonical $\rm S^{(o)} _{f} = 3\times 10^{-18}$ cm$ ^{3}$ s$ ^{-1}$ K$
^{-1/2}$ (Jura 1975a). The Cold  Neutral Medium (CNM) HI phase, out of
which  H$_2$ forms,  has typically  $\rm  T_k\sim (80-150)$  K with  a
slight  increase  with  galactocentric  distance $\rm  R_{gal}$  (e.g.
Braun 1997).   Using the  aforementioned H$_2$ formation  criterion or
equivalent expressions the steep $\rm H_2\rightarrow HI$ transition at
a particular $\rm  R_{tr}$ in spiral disks has  been modelled, and its
sensitivity  to variations  of pressure  and  metallicity demonstrated
(Elmegreen 1993;  Honma, Sofue, \& Arimoto 1995).   However, all these
studies have not examined the  effect of a larger H$_2$ formation rate
which will simply be to  ``push'' the molecular-rich part to encompass
a larger portion of a typical disk.

Molecular hydrogen  formation rates that  are $\sim 4-8$  times higher
than the standard value  are compatible with observations, since these
always contain an $\rm n-S_f$  ``degeneracy'' that can yield high $\rm
S_f$ values  if the gas is  less dense than assumed  (e.g.  Jura 1974;
Jura  1975a,b).    Results  from  {\it  ISO}   observations  of  H$_2$
rotational lines from photodissociation  regions (Habart et al.  2000;
Li et  al.  2002) strongly favor  $\mu = \rm  S_{f}/S^{(o)} _{f}\ga 5$
(see  also   Sternberg  1988  for  early  indications   of  such  high
values).  If this is  indeed the  case S  will be  raised at  all $\rm
R_{gal}$.  This rescaling has no  effect for those regions of the disk
where the old values are already  $\rm S>1$ and which CO imaging shows
them to  be indeed H$_2$-rich,  but now a  larger portion of  the disk
will have $\rm  S > 1$ and thus conditions  favorable for the presence
of a molecular gas phase.

The dependance of the spatial extent  of such a gas phase on the value
of the  formation rate can be  easily demonstrated from  Equation 1 if
the  FUV  volume emissivity  $\rm  j_{\circ  }$,  its dust  absorption
coefficient $\rm k_{d}$,  and the metallicity are assumed  to follow a
similar spatial distribution (Honma  et al.  1995).  Hence, neglecting
the  H$_2$  self-shielding   contributions  to  the  total  absorption
coefficient  it is  $\rm  z/G_{\circ}=z/(j_{\circ}/k_{d}) \propto  z$.
For an exponential disk profile  and defining $\rm S(R_{tr})\sim 1$ it
can be  shown that  the new $\rm  H_2\rightarrow H$  transition radius
will be

\begin{equation}
\rm R_{tr}(\mu )=R_{tr}(1)+(ln\mu) R_{e},
\end{equation}

\noindent
where $\rm R_{e}$ is  the scale-length of the exponential distribution
of optical  light and metallicity.   For $\mu =5$ the  molecular front
will now reside at least  $\rm (ln5) \rm R_{e}\sim 1.6 R_{e}$ further,
past that  inferred by CO imaging.   In typical disks the  later is at
$\rm  R_{tr}(1)=(0.5-1) R_{e}$  (e.g.  Young  \& Scoville  1991), thus
$\rm R_{tr}(\mu )\sim (2-2.5) R_{e}$, which places the molecular front
well inside a typical HI disk.  A flattening of the abundance gradient
and thus  of dust grain surface  per H nuclei at  large $\rm R_{gal}$,
and inclusion  of H$_2$ self-shielding strengthen  the above arguments
by  yielding  still  larger values  of  $\rm  S$  in the  outer  parts
of~disks.

\section{H$_2$ gas in metal-poor regions: CO-deficient, diffuse, and warm}

In  a metal-poor  H$_2$ gas  phase with  densities of  $\rm  n\sim 20\
cm^{-3}$ the lack  of the dissociated CO molecule  acting as a coolant
through its numerous rotational transitions causes the gas temperature
to be $\rm  T_k \sim 100$ K, similar  to that found for CNM  HI over a
wide  range of  galactocentric distances  (Papadopoulos, Thi,  \& Viti
2002). This is not surprising  since both the atomic and any molecular
gas phase  comprising CNM are  subjected to the same  dominant heating
(electrons ejected from  grains by FUV light) and  cooling (C$^+$ line
emission) mechanisms.   The additional H$_2$  cooling through emission
from its  quadrupole transition  S(0): $\rm J_{u}-J_{l}=2-0$  while it
does  not significantly  alter the  heating/cooling balance  away from
that of the atomic CNM phase,  it is still not negligible.  Indeed for
an H$_2$ gas  phase with volume densities and  temperatures typical of
the CNM, S(0) with its thermalization density of $\rm n_{20}\sim 54\rm
\ cm^{-3}$ and $\rm E_{20}/k\sim 510\ K$, is the only H$_2$ transition
that can be significantly excited.

This line is optically thin and offers a potent observational tool for
discovering whether the CNM in galaxies indeed harbors a molecular gas
phase and  can also yield an  estimate of its  mass.  Tantalizing S(0)
emission that may  be originating from such a  diffuse H$_2$ gas phase
has been detected with {\it ISO}  in the edge-on spiral galaxy NGC 891
whose  mass  may  outweighs that  of  HI  by  factors of  $\sim  5-15$
(Valentijn \& van der Werf  1999). Sensitive observations of S(0) line
emission  in a  variety of  metal-poor  environments hold  the key  in
revealing  whether such  a molecular  gas phase  exists and  with what
H$_2$/HI mass fraction.

\section{Conclusions}

The Cold  Neutral Medium  in galaxies may  contain an H$_2$  gas phase
with  similar  temperatures and  densities  that  can be  particularly
prominent  in their  metal-poor regions.   Its spatial  extent depends
sensitively on  the value of the  H$_2$ formation rate  on dust grains
and the  most promising  tool for revealing  its mass  distribution is
through the emission of  its optically thin S(0) $\rm J_{u}-J_{l}=2-0$
line.

\end{document}